\documentclass[acmtog, screen ,nonacm]{acmart}

\AtBeginDocument{%
  \providecommand\BibTeX{{%
    \normalfont B\kern-0.5em{\scshape i\kern-0.25em b}\kern-0.8em\TeX}}}

\setcopyright{none}
\copyrightyear{2022}
\acmYear{2022}
\acmDOI{XXXXXXX.XXXXXXX}

\usepackage{subfigure}
\usepackage{multirow}
\usepackage{xcolor}

\usepackage{algorithm,algcompatible,amsmath}
\algnewcommand\INPUT{\item[\textbf{Input:}]}%
\algnewcommand\OUTPUT{\item[\textbf{Output:}]}%

\algnewcommand{\algorithmicforeach}{\textbf{for each}}
\algdef{SE}[FOR]{ForEach}{EndForEach}[1]
  {\algorithmicforeach\ #1\ \algorithmicdo}
  {\algorithmicend\ \algorithmicforeach}

\begin{document}

\title[Progressive tearing and cutting of soft-bodies in high-performance virtual reality ]%
      {Progressive tearing and cutting of soft-bodies in high-performance virtual reality }

\author{Manos Kamarianakis}
\email{kamarianakis@uoc.gr}
\orcid{0000-0001-6577-0354}
\affiliation{%
  \institution{FORTH - ICS, University of Crete, ORamaVR}
  \country{Greece}
}

\author{Antonis Protopsaltis}
\orcid{0000-0002-5670-1151}
\email{aprotopsaltis@uowm.gr}
\affiliation{%
  \institution{University of Western Macedonia, ORamaVR}
  \country{Greece}
}

\author{Dimitris Angelis}
\orcid{0000-0003-2751-7790}
\email{dimitris.aggelis@oramavr.com}
\affiliation{%
  \institution{University of Crete, ORamaVR}
  \country{Greece}
}

\author{Michail Tamiolakis}
\orcid{0000-0002-9393-3138}
\email{michalis.tamiolakis@oramavr.com}
\affiliation{%
  \institution{University of Crete, ORamaVR}
  \country{Greece}
}

\author{George Papagiannakis}
\orcid{0000-0002-2977-9850}
\email{papagian@ics.forth.gr}
\affiliation{%
  \institution{FORTH - ICS, University of Crete, ORamaVR}
  \country{Greece}
}

\renewcommand{\shortauthors}{Kamarianakis, 
Protopsaltis, et al.}


\begin{teaserfigure}
\includegraphics[width=0.95\textwidth]{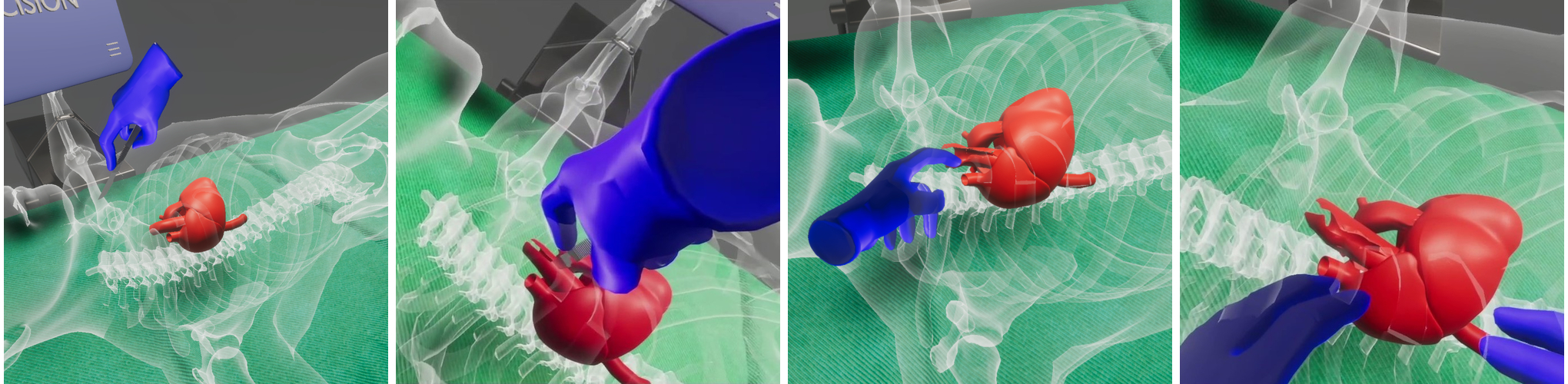}
\centering
\caption{A surgeon performing a highly-realistic tear operation on a pumping (animated/deformed while tearing) heart in VR. 
The operation is performed in real-time 10ms, 
while the 3D heart mesh is simulated and interacted as a soft-body. Further deformation on the soft tissues 
can be applied, being layered on top of a linear-blend skinned skeleton mesh.}
\label{fig:teaser}
\end{teaserfigure}

\begin{abstract}
   We  present an algorithm that allows a user within a virtual 
   environment
   to perform real-time unconstrained cuts or consecutive tears, i.e., progressive, continuous fractures on a 
   deformable rigged and soft-body mesh model in high-performance 10ms. 
   In order to recreate realistic results for different physically-principled materials 
   such as sponges, hard or soft tissues, we incorporate a novel
   soft-body deformation, via a particle 
   system layered on-top of a linear-blend skinning model. 
   Our framework allows the simulation of realistic, surgical-grade
   cuts and continuous tears, especially valuable in the context of medical VR 
   training. In order to achieve high performance in VR, our algorithms 
   are based on 
   Euclidean geometric predicates on the rigged 
   mesh, without requiring any specific model pre-processing. The 
   contribution of this work lies on the fact that current 
   frameworks supporting similar kinds of model tearing, 
   either do not operate in high-performance real-time or only apply to predefined
   tears. The framework presented allows the user to freely cut or
   tear a 3D mesh model in a consecutive way, under 10ms, while 
   preserving its soft-body behaviour and/or allowing further animation.
   
\end{abstract}  




\maketitle

\section{Introduction}

Since their inception, rigged animated models \cite{magnenat1988joint} 
have become a major research
topic in real-time computer graphics. 
Experts have been experimenting with 
various animation and deformation techniques, pushing the boundaries of realism and real-time performance. 
As the industry of Virtual, Augmented Reality (VR, AR) 
rapidly grows, the term of full user-immersion is being 
researched extensively. Fully-immersive virtual reality 
systems mainly aim to enable users to experience and 
perceive the virtual environments as real \cite{Protopsaltis2020}. 
To maintain user immersion at all times, these VR systems must 
produce and project a high number of frames per second, 
which implies that the computational latency for each frame 
should be minimal.
In this regard, increasingly more complex and 
optimized algorithms are being developed. 
Sophisticated computer graphics tools involve the ability to perform 
cuts, tears and drills on the surface of a skinned model
\cite{Bruyns:2002jc,Kamarianakis_Papagiannakis_2021}. 
Such algorithms are aiming to increase user immersion 
and to be used as sub-modules of even more complex operations. 
However, their scale-up for the extreme real-time conditions of 
virtual reality environments utilizing mobile, all-in-one 
un-tethered head-mounted displays (HMDs), remains an active 
field of research.

The need to interact in a shared virtual environment with other 
participants in the upcoming metaverse
pushes the envelope for more realistic deformation 
simulations that lead to more complex techniques and 
interaction paradigms. In the physical world, 
certain deformable objects, e.g., a soft or hard tissues, 
are deformed naturally when external forces are applied on them. 
To preserve immersion and avoid the uncanny valley in VR, the rigid object's 
physical behavior needs to be replicated in VR too
\cite{terzopoulos1987elastically,macklin2014unified,mages_poster}.
One way to accomplish this is via the so-called 
\emph{soft-body mesh deformation} \cite{mages3}, 
a suite of algorithms that essentially dictates how the vertices 
of a mesh should affect one another when 
an external force is applied anywhere on the surface of the model.

\begin{figure*}
    \centering
    \includegraphics[width=0.95\textwidth]{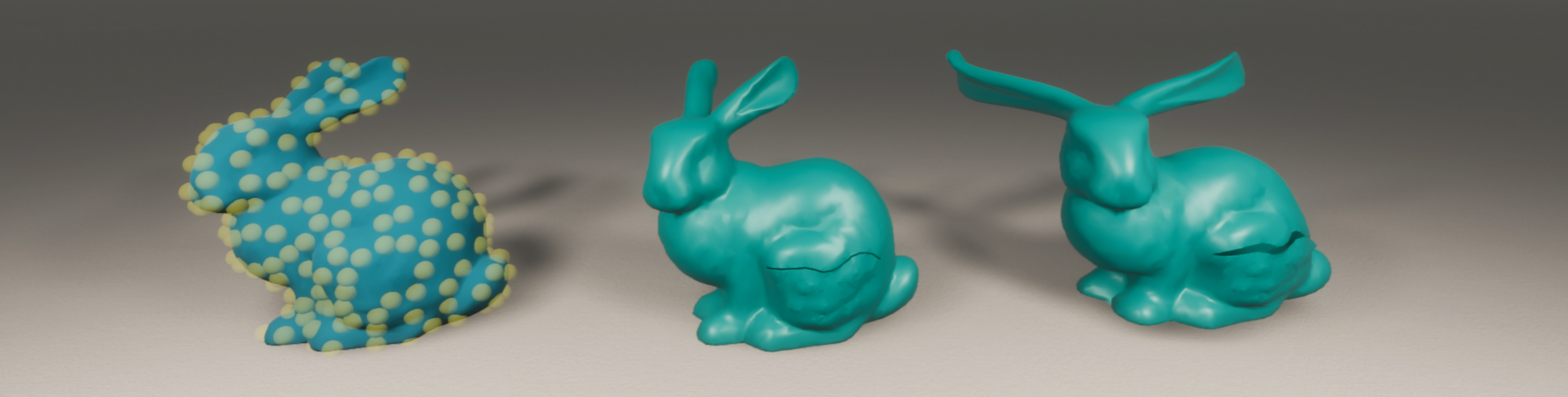}
    \caption{(Left) The vertices of a 3D  model are clustered in particles, to allow soft-body characteristics. (Middle) A continuous tear is 
  performed on the model. (Right) The particles of the torn model are updated, allowing further proper soft-body deformations.}
    \label{fig:bunnyTeaser}
\end{figure*}

Performing interactive cuts on a model is not something new; 
However, most techniques are not suitable for applications requiring 
high frame-rates as they are based on finite-element methods. 
Moreover, implemented cuts in such applications
are in most cases constrained: camera, model or user degrees of freedom, 
i.e. the user cannot freely cut anywhere on the model; a set of predefined 
cuts and their animations are usually produced and placed in the virtual environment  by 
VR designers or artists, and each one is played when triggered
by the user's specific and constrained actions.

In this work, we propose a framework that allows the user to
perform realistic tears, i.e., small cuts, on the surface of 
a model. Our algorithms are based on pure geometric operations 
on the surface mesh, and therefore are amenable to yield real-time 
results in VR, even in low-spec devices 
such as mobile VR head-mounted displays (HMDs). 
The significance of our work lies on the fact that in the current 
state-of-the-art, similar tears on a rigged 3D model in VR are 
predefined via linear-blend skinning animations, in order to allow them 
to playback in real-time. 
Our methods can be implemented in modern game engines such as Unity3D 
and Unreal Engine; convincing results are illustrated in the video 
accompanying this work (also, see Fig.~\ref{fig:teaser}).
The specific calculations must be performed in real-time 
within a 10-20 ms to preserve user immersion. 
The ongoing research for increased realism in virtual environments heavily impacts 
educational and training applications, especially the ones regarding 
VR medical training (and beyond)
\cite{papagiannakisEditorialNewVirtual2022}.



\section{Previous Related Work}

\cite{parkerRealtimeDeformationFracture2009} proposes a simplified version of previous FEM techniques for use in video-games and real-time simulations. They utilize a linearized semi-implicit solver and a well-mastered and optimized parallelized implementation on CPU of the conjugate gradient method. The adopted approach avoids re-meshing, by constraining the fracture on the faces of the simulation elements. It requires the duplication of vertices, while further introduces “splinters” that hide the produced artifacts. The embedded fracture model relies on maximum tensile stress criterion, element splitting according to a fracture plane, and local re-meshing to ensure a conforming mesh. This approach leads to a fast and robust fracture simulation for stiff and soft materials. 

\cite{mitchell2015gridiron} proposed a cutting algorithm, based on \cite{sifakisArbitraryCuttingDeformable2007}, that allows arbitrary cracks and incisions of tetrahedral deformable meshes. In their work, the utilization of low resolution meshes assists the efficient simulation of the model, while preserving the surface detail by embedding a high-resolution material boundary mesh, for rendering and collision handling. The method allows the accurate cutting of high-resolution embedded meshes, arbitrary cutting of existing cuts, and progressive cuttings during object deformation. The utilized algorithm is based the virtual node algorithm, that duplicates elements intersecting with the cutting geometry, rather than splitting them. The extended algorithm allows arbitrarily generalized cutting surfaces at smaller scales than tetrahedron resolution, and improves the shortcoming of the original algorithm, that restricted one cut per face and did not handle degenerate cases. The algorithm is based on embedding cracks in virtual elements, which limits the accuracy of the crack propagation computations. In this work several offline progressive cutting use cases were simulated by the proposed algorithm.

\begin{figure}[t]
    \centering
    \includegraphics[width=0.45\textwidth]{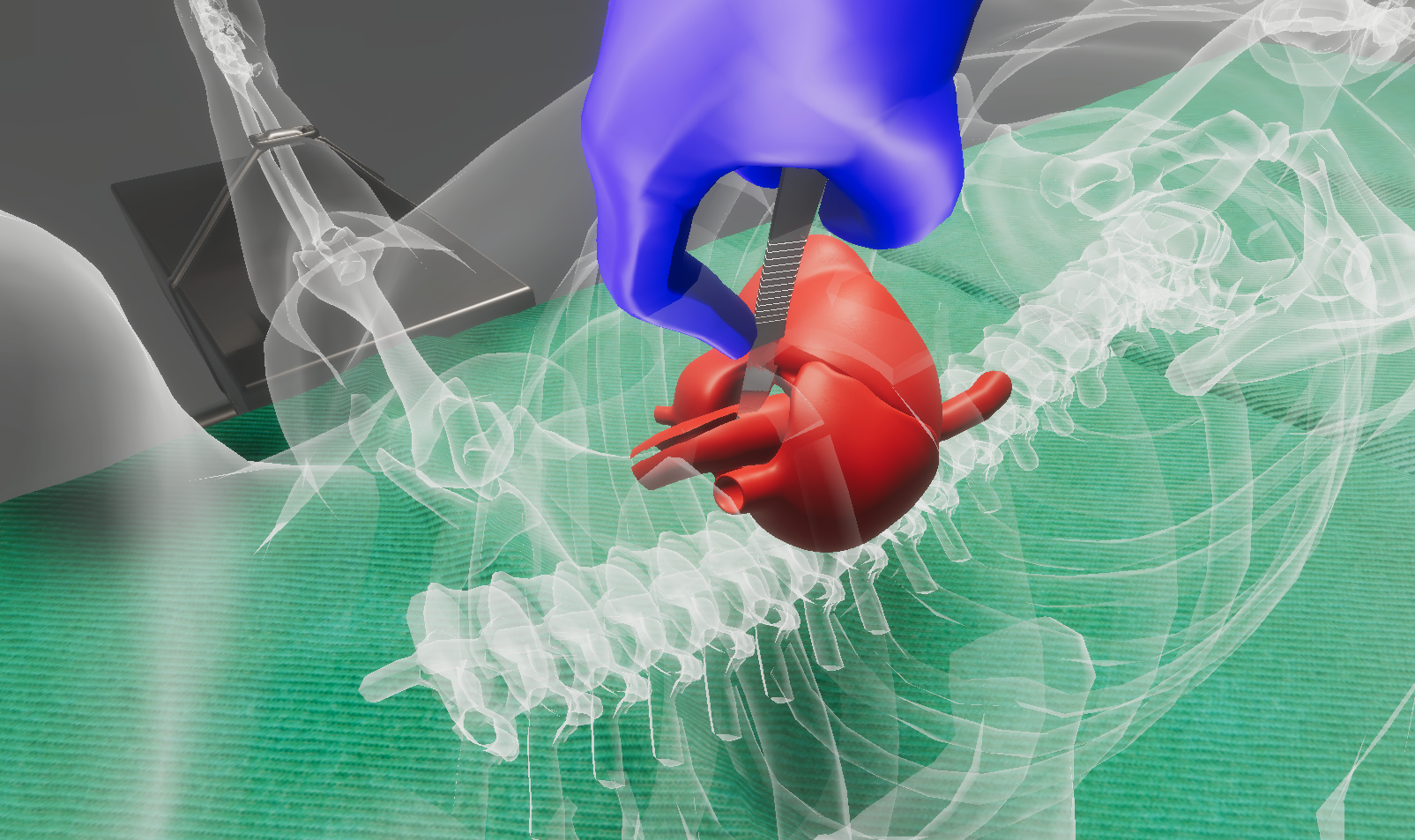}
    \caption{A close up of a deformable heart being torn by a scalpel. 
    The heart is simulated as a soft body using the proposed particle system.}
    \label{fig:heart_torn}
\end{figure}

Aiming to model physical object cutting behavior, \cite{heVersatileCuttingFracture2022} proposes an algorithm for highly realistic virtual cutting simulation, showing the contact effect before the cutting occurs, that considers deformable objects’ fracture resistance. It utilizes a versatile energy-based cutting fracture evolution model, based on Griffith’s energy. It introduces a tailored cut-incision evolution scheme that constraints the cutting tool's interaction with the deformable object, by evaluating the stage at which the cutting starts. To allow the surface indentation prior to cutting, the adapted model uses a material-aware scheme to generate the appropriate realistic and consistent behavior of the cutting tool, and the visual indentation deformation of the object. The designed framework is based on the co-rotational linear FEM model to support large deformations of soft objects and also adopt the composite finite element method (CFEM) to balance between simulation accuracy and efficiency. Additionally, it handles the collision and cut incorporation in the same way as the current FEM-based cutting methods using hexahedral elements. The experimental results show that realistic cutting simulations of various deformable objects with various materials and geometrical characteristics that introduce small computational cost for desktop systems.

Li et al. \cite{li2021interactive} propose real-time tearing and cutting operations on deformable surfaces using local Cholesky factorization updates in the global pass of a projective dynamics solver. These updates assist the handling of the simulations with topology changes. This adopted approach involves addition of new vertices, topological changes, and re-meshing operations, and allows effective gradual and progressive updates, which is common in real-time physics based simulations.

\begin{figure}
    \centering
    \includegraphics[width=0.45\textwidth]{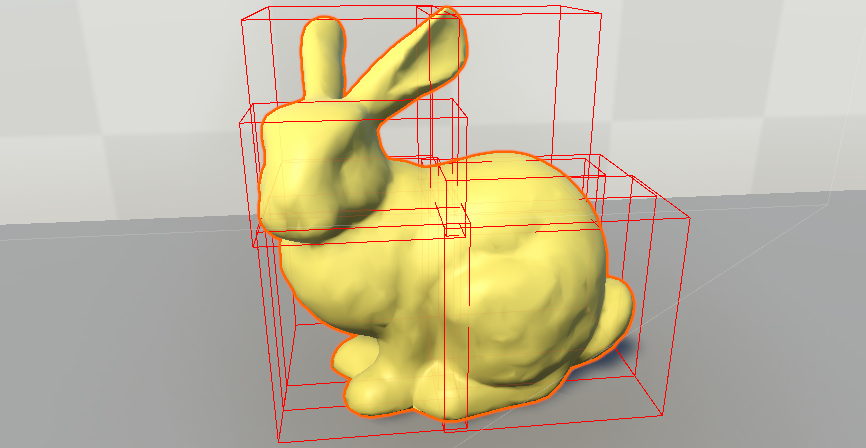}
    \caption{The segmentation of a bunny model into mesh-sections defined by axis-aligned bounding boxes. In this case, all boxes contain the same number of vertices. Mesh-section overlapping is required to handle tearings that involve faces contained in multiple sections.}
    \label{fig:bunnySegmented}
\end{figure}


\section{Our Approach}

Our methodology is based on the techniques of
\cite{Kamarianakis_Papagiannakis_2021}, where the authors
describe simple cut and tear operations on a
3D mesh using basic geometric operations. Our optimized
tearing module (see Section~\ref{sub:TheTearAlgorithm}) allow
for progressive uninterrupted
tears i.e., the user can freely perform tears successively
similar to a surgeon’s tearing gesture. Furthermore, our cutting module
(see Section~\ref{sub:TheCutAlgorithm}) performs straight cuts,
producing two (or more) fully deformable sub-meshes. Both modules
operate on deformable meshes, using geometric algebra operations.
To accomplish the so-called \emph{soft-body mesh deformation},
we have developed a suitable particle
decomposition on the model’s vertices based on \cite{nealen2006physically},
where the model’s vertices are clustered into groups,
and physics \emph{particles}
are assigned on each group to handle forces and collisions (see Section~\ref{sub:TheParticleSystem}).
Upon mesh import, the particles are generated as described in
Section \ref{ssub:GeneratingtheInitialParticles}, thus enabling
soft-body behaviour in the original model. The pipeline
used to properly simulate this behaviour in a modern game engine
is provided in Section~\ref{ssub:ParticleSimulation}.
After performing a tear (see Section~\ref{sub:TheTearAlgorithm})
or cut operation (see Section~\ref{sub:TheCutAlgorithm})
on the model, apart from the partial
re-meshing that the model undergoes, a subsequent update of
the nearby particles, involving affected vertices, is also required
(see Section~\ref{ssub:UpdatingtheParticlesAfterTear}).
This crucial step increases the realism of the torn model,
as it allows proper visual simulation, such as deforming or
animating, of the torn area.
Finally, in Section~\ref{ssub:AddingMoreParticles}, we propose an
optional step towards optimizing the visual outputs of a
torn soft-body model.

Via the proposed algorithms, we are able to perform real-time continuous
tears on a soft-body model and update the underlying particle
decomposition to obtain highly realistic results in VR.
Our methods were designed with the lowest possible computational complexity to yield real-time results and high frame-rates in VR.
Lastly, proper handling and weight assignment
\cite{Kamarianakis_Papagiannakis_2021} to the tear-generated vertices
allow us to tear not only rigid but also skinned models, where in the
latter case, further animation is still feasible.

\subsection{The Tear Algorithm}
\label{sub:TheTearAlgorithm}

In order to achieve real-time tearing results, we have opted for basic geometric
primitives, e.g., face-plane intersections and face ray-casting, as 
basic building blocks for our algorithms. 
This  approach allows for fast  identification of the faces affected 
by the tear. 

In our implementation, the tear width is user defined. In non-zero settings, 
a destructive tear takes place: faces that fall in the tear-gap are 
completely or partially \emph{clipped}, i.e., removed from the model. 
Partially clipped faces are calculated by their intersections with the 
tear-gap surrounding box which is defined by 
``connecting'' consecutive  bounding-boxes of single 
tear segments.

\begin{figure}
    \centering
    \includegraphics[width=0.45\textwidth]{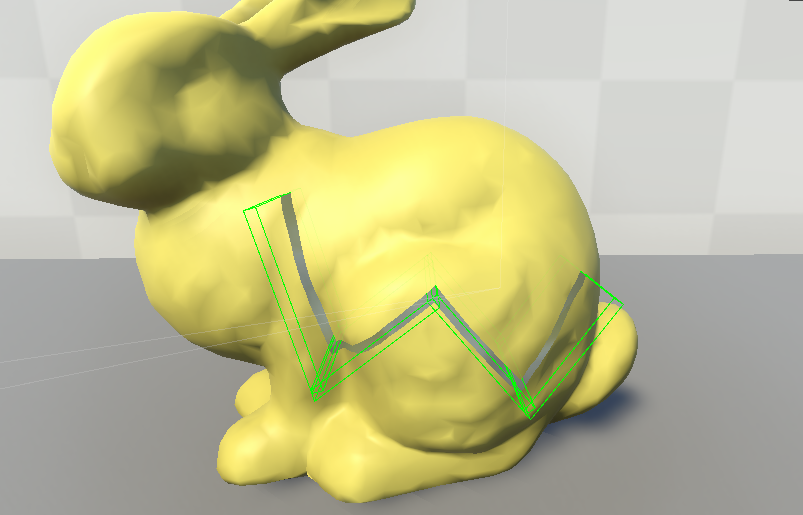}\\
    \includegraphics[width=0.45\textwidth]{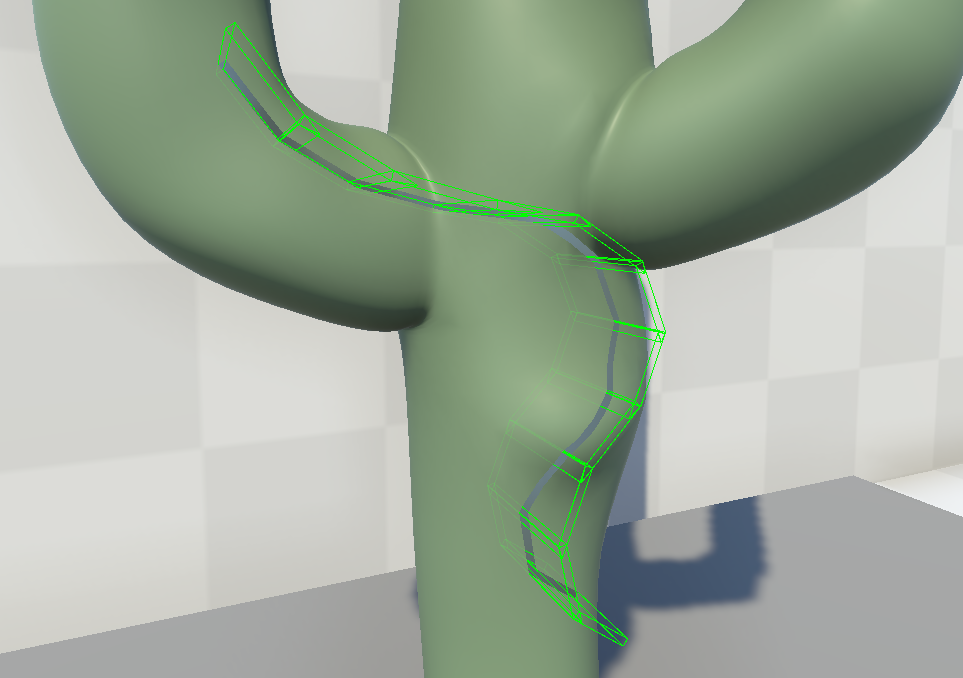}
    \caption{Bounding boxes defined during a tear. The scalpel's position 
    is sampled based on a time or distance threshold, and a sequence of
    over overlapping bounding boxes is created. A larger threshold is used in 
    the bunny model, whereas a smaller one on the cactus; the length of the 
    boxes changes proportionally to this threshold.}
    \label{fig:boundingBoxes}
\end{figure}

In case of a single tear segment, such a bounding box 
is aligned and 
bounded by the scalpel's endpoints in its
final position and the scalpel's intersection with the model in its initial position; the width of the box is equal to the user 
defined tear width (see Fig.~\ref{fig:boundingBoxes}).
As the user moves the scalpel, freely tearing the model,
several scalpel's positions are sampled at specific time or distance intervals, defining multiple consecutive tears segments. 
In case of abrupt movements in the scalpel's trajectory, the algorithm forces extra sampling on the scalpel's position.
To avoid jagged edges on the tearing path,
the algorithm makes sure that consecutive bounding boxes do not overlap, by utilizing  
non rectangular bounding boxes instead (see Fig.~\ref{fig:boundingBoxes}). 

To further optimize the performance of the tear algorithm, 
the mesh is segmented into smaller groups, called \emph{mesh sections}. 
Each mesh section is defined as an axis aligned bounding box and 
contains groups of the mesh faces. 
This division of the mesh into smaller sections 
reduces significantly the Tear algorithm running 
time, as the 
affected mesh section is only a small subset 
of the entire mesh (see Fig.~\ref{fig:bunnySegmented}).
The number and size of these sections are user defined.

Some comments on the steps of the tear algorithm are found 
below. 

\begin{algorithm}
\caption{Tearing Algorithm}
\label{alg:tearing}
\begin{algorithmic}[1]
\INPUT Triangulated Mesh $M$, scalpel's position 
at specific time steps, Mesh  Sections. 
\REQUIRE Scalpel properly intersects $M$ at these time steps
and that a tearing plane between timesteps are properly defined.
\OUTPUT The mesh resulting from $M$ getting torn
    by the scalpel.
\ForEach{Two consecutive timesteps}: 
    \State Define a bounding box for the scalpel's tear segment.
    \State Smoothen out the intersection of two bounding 
    boxes by replacing a plane of the leading box by a 
    the ``touching'' of the next one.
    \State 
    Determine the vertices of the mesh section(s) that fall into the 
    bounding boxes, by performing multiple side tests 
    of the planes defining the bounding boxes.
    \State 
    Define a search list $S$ containing all faces 
    containing such vertices. \label{alg:creatingS}
    \ForEach{face $T$ in $S$}\label{alg:triangleLoop}
        \ForEach{plane $\Pi$ of the first bounding box $B$}
            \IF{$T$ intersects $\Pi$}
                \IF{The intersection points fall inside the band of the neighbouring planes of $\Pi$} \label{alg:intersectionTest}
                \State Retriangulate the face into two smaller ones.
                \State Keep only the smaller face(s) that lie outside $B$, thus clipping the mesh inside $B$.
                \State Determine the normals/uvs of the intersection points via interpolation. 
                In case of a rigged model, also determine the intersection point weight values from the nearby vertices. \label{alg:interpolation}
                \ENDIF
            \ENDIF
        \EndForEach    
    \EndForEach
    \ForEach{Smaller face kept from previous loop}
    \State Do a second retriangulation pass, i.e., search its neighboring face and split it into two parts, if not split already. \label{alg:secondPass}
    \EndForEach
\EndForEach
    \State Update the mesh model and finalize the tear, by sending 
the cached final vertices and faces to the GPU buffer 
to properly update the mesh model.
  \end{algorithmic}
  \end{algorithm}

\begin{figure}
    \centering
    \includegraphics[height=4cm,width=0.22\textwidth]{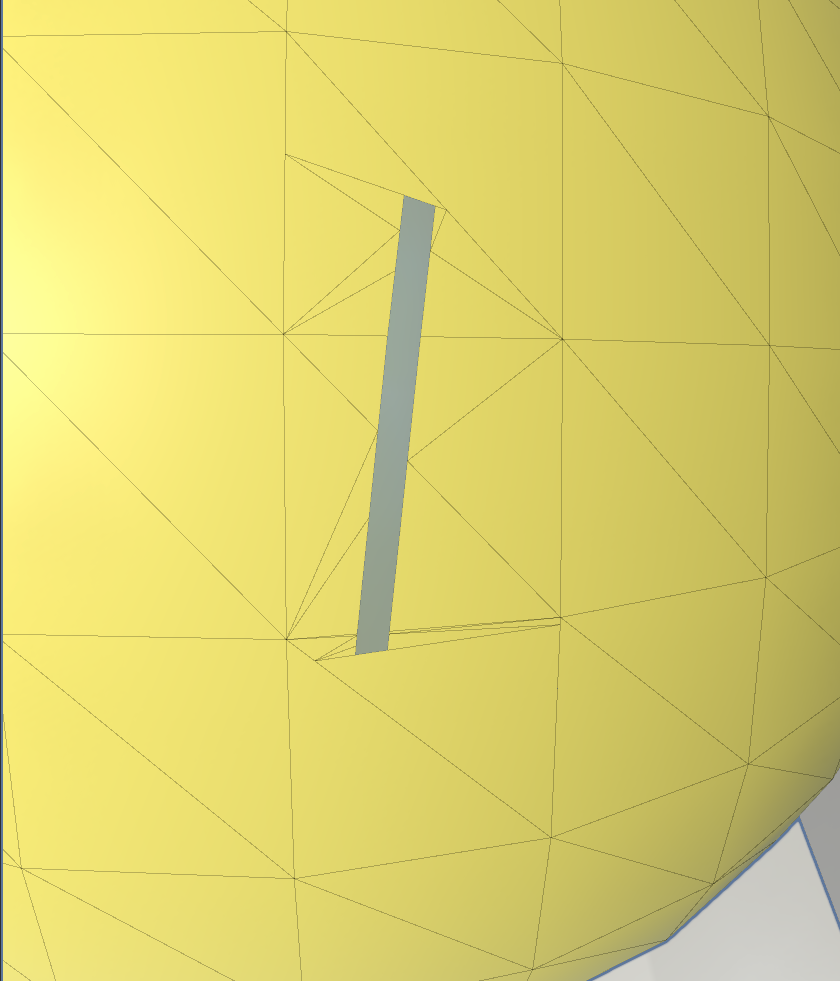}
    \includegraphics[height=4cm,width=0.22\textwidth]{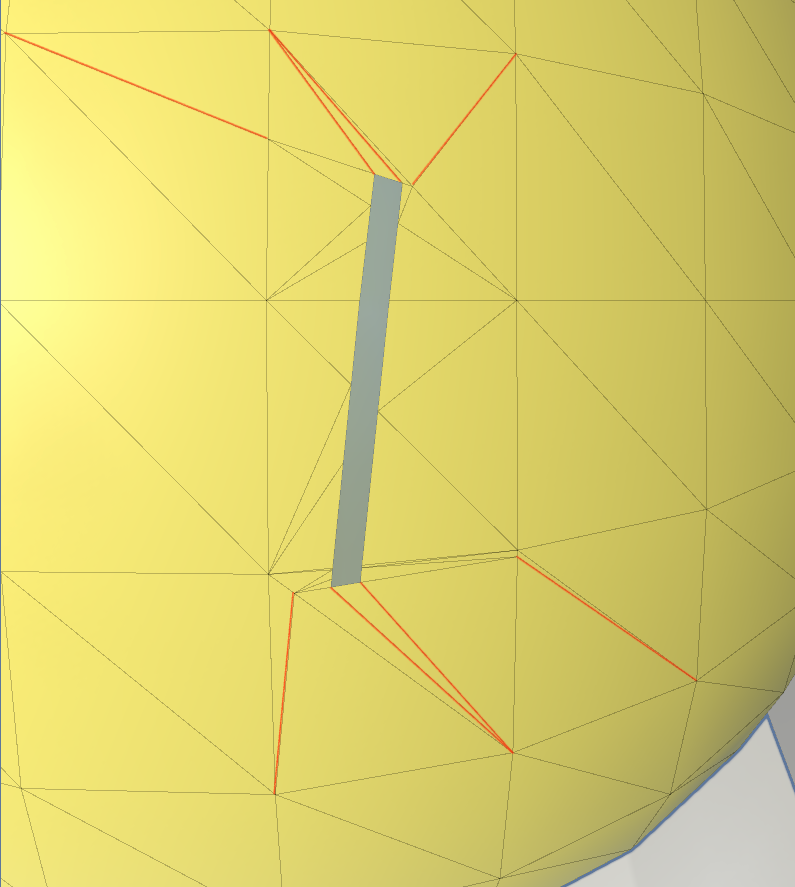}
    \caption{An additional triangulation pass ensures an intersection 
    point on an edge is properly connected to both vertices on the adjacent 
    faces. Left figure depicts the result before this ``second pass''; right
    figure shows, in red, the edges added in this step.}
    \label{fig:2ndPass}
\end{figure}

\begin{itemize}
\item 
In Line~\ref{alg:creatingS}, the 
affected mesh sections are identified and searched for  
faces to be added to the  search list $S$. 
This reduces running times, especially in complex models 
with a large number of vertices.
\item
In Line~\ref{alg:intersectionTest}, a more sophisticated 
check that ensures a tested face
$T$ will remain unaffected in the torn model, 
and therefore, it can be removed from $S$ for the 
next iteration, which further reduces the loop iterations in
Line~\ref{alg:triangleLoop}.
\item
The operation in Line~\ref{alg:secondPass}, ensures 
that a  properly triangulated mesh will be produced, 
i.e., a vertex on an edge will be connected to 
the opposite vertex in both adjacent faces
(see Fig.~\ref{fig:2ndPass}).
\item 
In Line~\ref{alg:interpolation}, if the model is 
a soft-body (see Section~\ref{sub:TheParticleSystem}), 
the particles map is also updated (see Section~\ref{ssub:UpdatingtheParticlesAfterTear}).
\end{itemize}

\subsection{The Cut Algorithm}
\label{sub:TheCutAlgorithm}

The algorithm to perform a thorough straight cut on the 
mesh model is a
simplified version of the respective
single tear algorithm.
Indeed, we define the cutting plane $\Pi$ as the plane that goes 
through the following three points:
the initial intersection point of the model mesh with the scalpel at a time step,
and the scalpel's endpoints after a specific time step;  
notice that these three points should  not be co-linear, 
otherwise the selected time step is altered.

\begin{algorithm}
\caption{Cutting Algorithm}
\label{alg:cutting}
\begin{algorithmic}[1]
\INPUT Triangulated Mesh $M$, cutting plane $\Pi$. 
\OUTPUT Two sub-meshes resulting from $M$ getting cut
    by the plane.
\ForEach{Face $T$ in $M$}
    \IF{$T$ intersects $\Pi$}
        \State Evaluate intersection points and add them to 
        the $M$.
        \State Determine the normals/uvs of the intersection point via interpolation. In case of a rigged model, also determine the intersection point weight values from the nearby vertices. \label{alg:cutInterpolation}
        \State Retriangulate $T$ into three smaller ones, 
        containing the intersection points. 
        \State Replace $T$ in the model with the smaller sub-faces.
    \ENDIF
\EndForEach
\ForEach{Face $T$ in $M$}
    \IF{Any vertex of $T$ lies on the positive side of $\Pi$}
    \State 
        Put $T$ and its vertices on the positive sub-mesh.
    \ELSIF{Any vertex of $T$ lies on the positive side of $\Pi$}
        \State 
        Put $T$ and its vertices on the negative sub-mesh.
    \ENDIF
\EndForEach
\end{algorithmic}
\end{algorithm}

\begin{figure}
    \centering
    \includegraphics[width=0.45\textwidth]{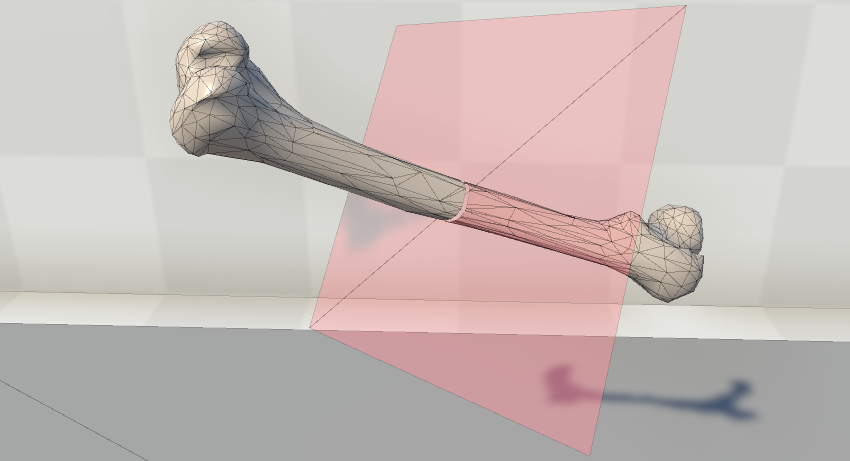}
    \caption{A bone is cut. The cutting plane, defined by the scalpel, 
    is denoted in red. The original bone model is dissected in two parts,
    each containing vertices lying on the same side of the plane.}
    \label{fig:FemurCutWhole}
\end{figure}

As in tear algorithm, if the model is 
a soft-body (see Section~\ref{sub:TheParticleSystem}), the particles map 
is also updated (see Section~\ref{ssub:UpdatingtheParticlesAfterTear}) 
during Line~\ref{alg:cutInterpolation}.
After applying the cutting algorithm, each 
sub-model will lie on the same side of the 
cutting plane (see Fig~\ref{fig:FemurCutWhole}).

\subsection{The Particle System}
\label{sub:TheParticleSystem}

\subsubsection{Generating the Initial Particles}
\label{ssub:GeneratingtheInitialParticles}

The offline process followed to 
generate the initial particles is summarized in 
Algorithm~\ref{alg:particleGeneration}.

\begin{algorithm}
\caption{Particle Generation Algorithm}
\label{alg:particleGeneration}
\begin{algorithmic}[1]
\INPUT Triangulated Mesh $M$, user defined particle radius $d$ 
and particle-to-particle distance $\delta$.
\OUTPUT Particle map for $M$.
\State
Perform a Poisson Disk Sampling, on the vertices of $M$. 
\label{alg:PoissonSampling}
\State
Spawn spherical particles centered at the selected vertices, which are denoted as \emph{anchor points}. \label{alg:particleSpawn}
\State
The particle and its anchor are connected via a spring.
\label{alg:springConnection}
\ForEach{Particle $P$}
    \ForEach{Vertex $V$ of the mesh}
        \IF{$V$ lies inside the sphere, centered 
        at $P$'s anchor point, with radius $d$.} 
        \State Assign the vertex $V$ to the particle $P$ and determine the weight influence of $P$ on $V$. \label{alg:particlesAssignVertices}
        \ENDIF
    \EndForEach
    \ForEach{Other particle $Q$}
        \IF{$Q$'s anchor point lies inside the sphere, centered 
        at $P$'s anchor point, with radius $\delta$.} 
        \State Denote $Q$ and $P$ as neighbouring 
        particles and determine their in-between influence 
        \label{alg:particlesNeighbours}
        \ENDIF
    \EndForEach
\EndForEach
\end{algorithmic}
\end{algorithm}

\begin{figure}
    \centering
    \includegraphics[width=0.47\textwidth]{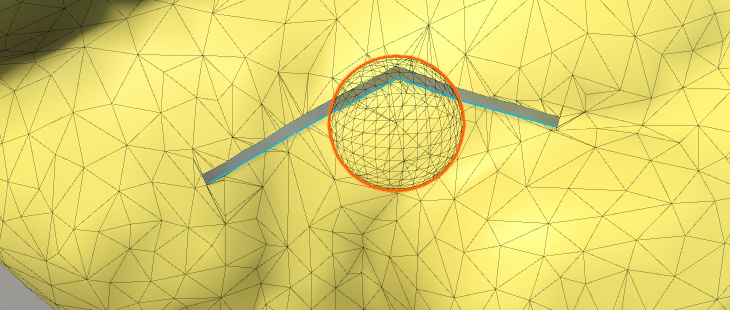}\\
    \includegraphics[width=0.47\textwidth]{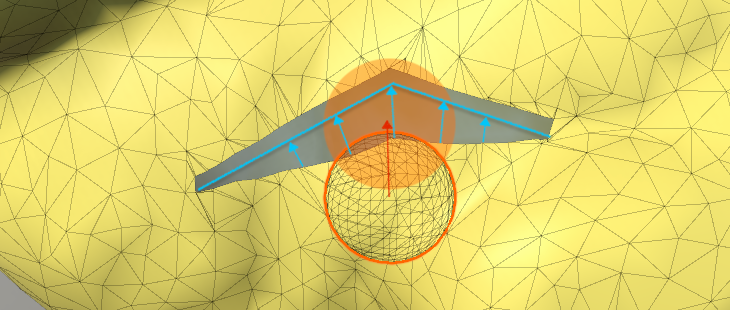}\\
    \caption{As a particle (orange circle) is displaced from its initial 
    position(Top), it gains a velocity that tends to return it there (red arrow), 
    simulating elasticity on all affected vertices(Bottom). Blue arrows represent 
    the resulting tendency of the vertices 
    to return to their initial position.}
    \label{fig:particleDisplacement}
\end{figure}

\begin{figure}
    \centering
    \includegraphics[width=0.47\textwidth]{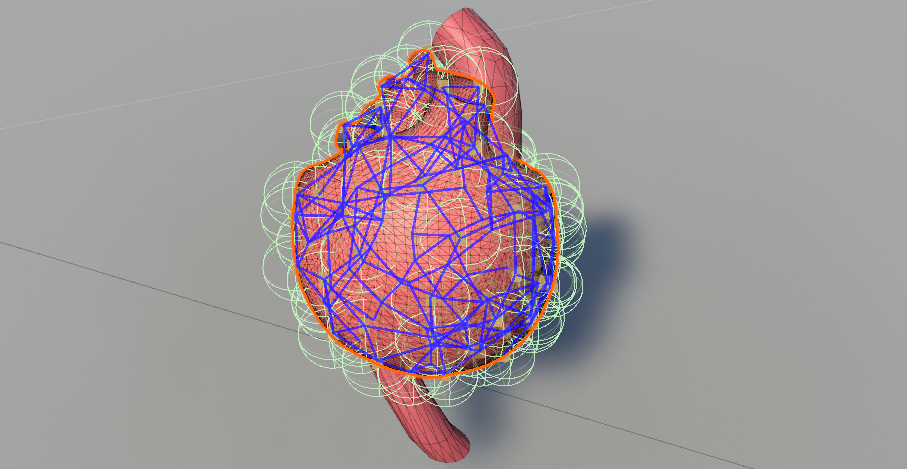}
    \caption{The particle system layer on top of a heart model. 
    The spheres illustrate the  particles and their interconnections (in blue lines).}
    \label{fig:BunnyParticleConnections}
\end{figure}

Some remarks on the particle generation algorithm can 
be found below
\begin{itemize}
\item 
The Poisson Disk Sampling \cite{bowersParallelPoissonDisk2010} in Line~\ref{alg:PoissonSampling}
ensures that the selected points 
will be uniformly distributed on the mesh,
maintaining sufficient distance between them.
\item 
In terms of modern game engines 
(e.g., Unity3D), we would describe the 
particles spawned in Line~\ref{alg:particleSpawn} as a GameObject with a spherical
collider and a rigidbody to handle physics forces. 
\item 
To properly simulate soft-bodies, 
we apply a typical spring mass approach 
\cite{nealen2006physically}, 
with some modifications (e.g. the inside pressure is not used to calculate the 
movement of each particle). 
In Line~\ref{alg:springConnection}, each particle gets connected to its anchor point's 
position vertex, via a spring, thus ensuring that the 
particles will always tend to return to their initial anchor 
position upon displacement. (see Fig.~\ref{fig:particleDisplacement})
\item 
In Line~\ref{alg:particlesAssignVertices}, 
the vertices assigned
to the particle $P$ are the ones that will be affected by $P$'s
potential displacement, with a weight inversely 
proportional to their distance from the $P$'s spawn position, 
usually based on a sigmoid function. 
\item 
In Line~\ref{alg:particlesNeighbours}
we may determine each particle's adjacent particle 
neighbours. 
This grid of particles (see Fig~\ref{fig:BunnyParticleConnections}) will eventually act as a set of
control points that will enable soft-body deformations; upon moving a particle, 
all neighbouring particles will also be partially displaced, and the 
affected vertices will yield the desired effect. 
A particle's displacement due to the 
movement of an adjacent particle is inversely 
proportional to the distance of their anchor points.
\end{itemize}

The connections between particles and vertices 
affected or neighbouring particles,
along with the respective influence weights, 
are referred to as  the \emph{ particle  map.}

\subsubsection{Particle Simulation} 
\label{ssub:ParticleSimulation}

The particle simulation is performed almost natively 
by modern game engines, such as Unity3D, as it involves 
the same mechanics with joint animation, i.e., 
both frameworks include control points, and 
vertices with weights assigned to them.

Particularly, for static meshes, only the particles' anchor 
positions need to be updated. 
In each simulation update, Unity3D automatically 
calculates forces and collisions, and applies the 
position changes for each particle, and 
the corresponding weighted displacement on 
the assigned vertices. 
On the other hand, skinned meshes involve additional steps in each simulation update. 
Initially, the particle's anchor position is calculated based on the pose of the 
model at the specific time step. As the anchor points 
are essentially vertices of the mesh, 
the animation equation is applied to obtain these positions. 
Subsequently the particle's adjacency is updated, by re-applying 
Line~4 of Section~\ref{ssub:GeneratingtheInitialParticles}; 
as the position of anchor points might have been altered, 
the particles' adjacent neighbours may have changed, based 
on the $\delta$ distance threshold. Finally, 
the same mechanics with a static mesh are applied, i.e.,
determine the position of the particle and consequently of 
the affected vertices. Specifically, 
the final global position $f_i$ of the $i$-th vertex of the 
model is determined, by evaluating

\begin{equation}
\label{eq:particleEquation}
f_i = \sum_{j\in J(i)} w_{i,j}D_j v_i,
\end{equation}

where 
\begin{itemize}
\item 
$f_i$ and $v_i$ are the final and initial homogeneous coordinates 
of the $i$-th vertex,
\item $J(i)$ contains the indices of the particles that affect 
the $i$-th vertex,
\item  $w_{i,j}$ is the corresponding influence factor between
the $j$-th particle and the $i$-th vertex, and
\item $D_j$ is the 4x4 matrix corresponding to the 
displacement of the $j$-th particle from its anchor, in global 
coordinates.
\end{itemize}

We obtain the final displacement $D_j$, either directly, i.e., the user 
moved a specific set of particles, or indirectly, via model-user interaction 
in VR, e.g., the user squeezed the model. 
In the latter case, the game engine's physics component 
is responsible to evaluate the displacement $D_j$ of the particles, by 
calculating the forces and collisions involved, at runtime. 
Under the physics engine hood, 
the forces applied to particles displace them from 
their anchor points. As a consequence, their velocity is altered to 
be proportional to the respective displacement, always pointing to the initial 
position, thus simulating elasticity. 

In the former case, where the user chooses to displace a specific 
set of particles, we may evaluate the displacement of all particles, 
by taking into consideration that the particles are interconnected via 
a spring-like system. Thus, we may determine the displacement $D_j$
of a particle indirectly, moved by its neighbouring particles movements
via 

\begin{equation}
\label{eq:particleEquationDisplacement}
D_j = \sum_{k\in K(j)} W_{j,k}D'_k,
\end{equation}

where 
\begin{itemize}
\item $K(j)$ contains the indices of the particles that are 
connected to the $j$-th vertex,
\item 
$D_j$ is the final displacement of the $j$-the particle, which was not 
displaced directly, but indirectly, due to the adjacency with 
the $k$-th vertex,
\item $D'_k$ is the 4x4 matrix corresponding to the 
displacement of the $k$-th particle that is displaced by the user
directly,
\item  $W_{j,k}$ is the corresponding influence factor between
the $j$-th and the $k$-th particle, a value inversely proportional
to their in-between distance.
\end{itemize}

\subsubsection{Updating the Particles After a Tear or a Cut}
\label{ssub:UpdatingtheParticlesAfterTear}

\begin{figure}
    \centering
    \includegraphics[width=0.45\textwidth]{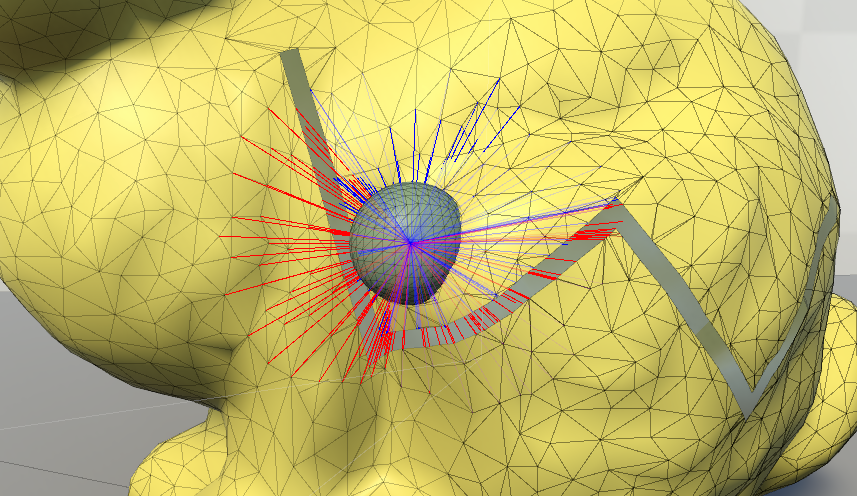}
    \caption{A spherical particle's (gray) anchor point (i.e., center) 
    is connected with segments to all initially affected vertices, before the tear operation. A red segment indicates that the corresponding vertex is 
    no longer affected after the tear, as a tear bounding box intersects with it. A blue 
    segment, indicates that the vertex is still affected after the tear.}
    \label{fig:particle_update}
\end{figure}

After a tear or a cut operation, it is 
important to update the particle map, 
in order to preserve the realism of the soft-bodies.
This map is updated by adding or removing vertices,
as well as modifying the particle connections 
with the assigned vertices or other 
neighbouring particles. 
To produce physically correct
deformation results, simple directives are introduced, e.g., 
vertices belonging to opposite sides of a tear, 
although close enough, cannot belong to the same particle
(see Fig.~\ref{fig:particle_update}). 

Below we provide an overview of the algorithm used to perform the particle update during a tear operation.

\begin{algorithm}
\caption{Particle Update Algorithm}
\label{alg:particleUpdate}
\begin{algorithmic}[1]
\State
Assign intersection points introduced by the 
tear operation to particles, as in Line~\ref{alg:particlesAssignVertices} of the Particle 
Generation Algorithm.
\State
If any tear bounding box intersects the segment 
connecting two neighbour particles, these particles 
are no longer considered neighbours.
\ForEach{Particle $P$}
\State
Remove any vertex $V$ affected by $P$ from its 
influence list if $V$ and the anchor point of $P$ lie 
on different sides of any tear plane. 
\label{alg:particlesVerticesDisconnection}
\EndForEach
\end{algorithmic}
\end{algorithm}

Regarding Line~\ref{alg:particlesVerticesDisconnection}, 
this simple-to-describe objective is one of the most 
important and challenging  primitives, in order to avoid 
potential artifacts. Special focus was given 
to intersection vertices that were introduced 
close to the connection of two consecutive 
tears segments, i.e., in the 
intersection of two bounding boxes. To properly identify 
whether a vertex lies on the opposite side of the anchor point 
with respect to a \emph{tear plane}, i.e., the plane splitting the bounding box in half, containing the scalpel's endpoints, 
we had to consider both the current and the previous tear 
segment. Additionally, to provide correct results in lower 
running times, for a given particle 
$P$, we only considered checking against tear planes 
that corresponded to sufficiently close bounding boxes. 
To be more specific, if the vertex, lying on a bounding box
(i.e., an intersection point introduced during the tear 
operation), closest to the anchor point of $P$, is not affected
by $P$, then the corresponding check of the particle against the corresponding tear plane may be omitted.

A similar but simpler methodology 
is followed after a cut operation on the mesh. 
In that case, the particle clustering map is updated
by removing all vertex-particle or particle-particle 
connections where the corresponding connection 
segment intersects the cutting plane.

\subsubsection{Adding More Particles for Optimized Tear Animation}
\label{ssub:AddingMoreParticles}

\begin{figure}
    \centering
    \includegraphics[width=0.47\textwidth]{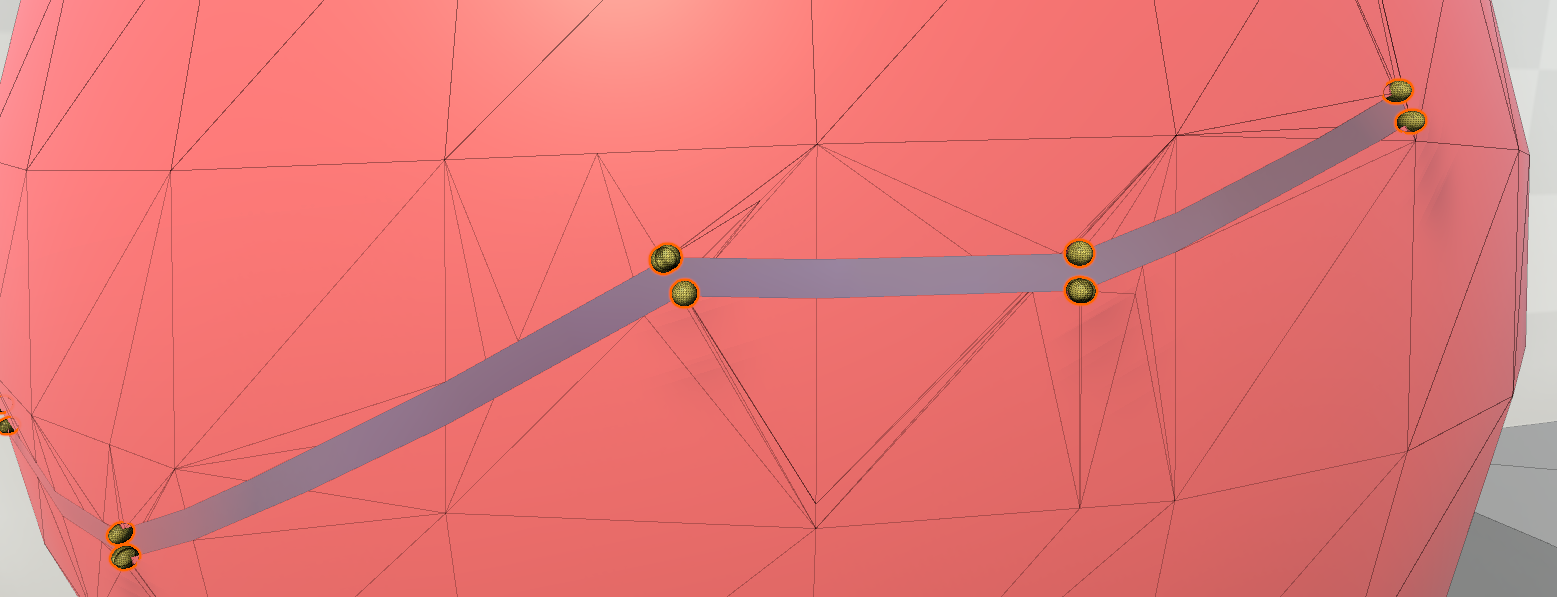}
    \caption{Location of particles (yellow spheres) additionally 
    spawned during a progressive tear towards improving visual results. 
    A movement of these particles away from the tear achieves higher 
    realism and a smoother animation of the tear opening up.}
    \label{fig:additionalParticles}
\end{figure}

The method proposed for  a progressive tear operation
on a mesh model with a subsequent update of the existing particles, 
yields highly realistic results.
However, the described particle system still misses to 
model the absolute physical behavior of human tissue. 
To model such realistic behavior, the reaction of human tissue 
after a tear operation would be to animate and slightly open up the wound. 
To achieve this type of animation, we consider an auxiliary set of 
newly created particles around the tear slit 
(see Fig.~\ref{fig:additionalParticles}). 
These new particles will be assigned to all vertices that 
took part in the two triangulation passes. 
A slight displacement of the new particles' anchors in a 
direction normal but away from the tear segments makes the 
animation possible.

However, the insertion of additional particles leads not 
only to increased realism, but also increased running times.

\section{Results and Discussion} \label{sec:results}

We experimented our method using various mesh models;
some representative use cases are shown in Figs.~\ref{fig:teaser}, 
\ref{fig:bunnyTeaser} and \ref{fig:heart_torn}. In all cases, further cut/tear operations, soft-body deformations 
and/or model animations are possible.
Tables~\ref{tbl:tear_running_times} and \ref{tbl:cut_running_times} 
contain the time required to 
perform the algorithms for tear and cut respectively.
All running times were obtained using a Windows 11 PC 
equipped with AMD Ryzen 7 5800H at 3.2GHZ, 16GB RAM and 
an Nvidia RTX 3060 (6GB RAM) graphics card.
The proposed method was implemented exclusively using non-parallel CPU computations. 
Our methods are also partially implemented in a VR medical 
training application \cite{CVRSB2021}, running on a modern
game engine performing identical results and producing real-time 
frame-rates,  suitable for desktop but also for VR immersive systems. 
In our experiments, an HTC Vive Pro tethered HMD was used to 
properly to validate that a satisfying immersive quality of experience 
(QoE) on the user's end was achievable using our framework; this is 
illustrated in the video accompanying this work.

\begin{table}
  \caption{Running times required to tear a sphere, a bunny 
  and a heart model.}
  \begin{center}
  \begin{tabular}{|c|c|c|c|}
  \hline
  Characteristics & Sphere & Bunny & Heart \\
  \hline
  \hline 
  Number of vertices & 515 & 2527 & 9747\\
  Number of faces & 768 & 4968 & 18336\\
  Number of particles & 191 & 179 & 224 \\
  \hline
  \hline 
  Operation & \multicolumn{3}{c|}{Running times per tear segment} \\
  \hline
  \hline
  Perform Tear & 0.36 ms & 3 ms & 2.54 ms \\ 
  Update particles & 0.39 ms  & 2.01 ms & 0.87 ms \\
  Disconnect Particles & 0.91 ms & 1.25 ms & 2.63 ms \\
  Calculate BoneWeights & 0.90 ms & 3.81 ms & 11.04 ms \\
  Update Mesh & 0.07 ms & 0.24 ms & 0.76 ms \\
  \hline
  \hline 
  Total Time & 3.25 ms & 11.19 ms & 18.65 ms \\
  \hline
  \end{tabular}
  \label{tbl:tear_running_times}
  \end{center}
\end{table}

Experiments of \cite{heVersatileCuttingFracture2022} showed partial cutting simulations of various deformable objects, called fractures, which correspond to our tear operations. The method produces highly realistic virtual cutting simulations considering the deformable object's fracture resistance. The scalpel cutting seems to be a little inaccurate with respect to the visualized cut. In terms of computational results, the method introduces a relatively low overhead on desktop stations while no experimentation is mentioned on demanding frame-rate systems such as VR or embedded within game engine pipelines. The experiments show (see Table~\ref{tbl:tear_comparison}) that for medium sized models the method’s results are close to our results, while in larger models our method is much faster.

The work of \cite{mitchell2015gridiron} provide cutting results for various models with no information on their mesh resolution. The experimentation of the method was run on two different desktop platforms. The running times of the method are bound by the utilization of a Newton-Raphson iterative solution scheme. In that regard, the method was not experimented in VR or game engine environments.

The \cite{li2021interactive} provide real-time tearing and cutting operations on deformable  surfaces. This method is mainly experimented on cloth models which differ significantly from  surgical-like tear operations. The simulation involves a local/global solve of projective dynamics with the pre-computed factorization, and the factor modification process. The produced results are indeed satisfying for desktop systems, but not for VR, as the time for a cloth cut is 49ms in total.

\begin{table}
  \caption{Comparison of ours tear method with the one
  presented in \cite{heVersatileCuttingFracture2022}.  
  Models of similar complexity were used. 
  OUR denotes that the proposed framework was applied. 
  }
  \begin{center}
  \begin{tabular}{|c|c|c|}
  \hline
  Model & Faces & Running Time\\
  \hline
  \hline 
Horse & 4266 & 10.14 ms\\
Bunny (OUR) & 4968 & 11.19 ms\\
\hline
Cuboid & 18128 & 52.77 ms\\
Heart (OUR) & 18336 & 18.65 ms\\
  \hline 
  \end{tabular}
  \label{tbl:tear_comparison}
  \end{center}
\end{table}

\begin{table}
  \caption{Running times required to cut a bone, a bunny 
  and a small cactus model.}
  \begin{center}
  \begin{tabular}{|c|c|c|c|}
  \hline
  Characteristics & Bone &  Bunny & Small Cactus \\
  \hline
  \hline 
  Number of vertices & 516 &  2527 & 2976\\
  Number of faces & 983  & 4968 & 3000\\
  \hline
  \hline 
  \multicolumn{4}{|c|}{Cut Operation} \\
  \hline
  \hline
  Intersection Points & 64  & 356 & 186 \\
  \hline
  Running times & 12 ms  & 17.29 ms & 13.49 ms \\ 
  \hline
  \end{tabular}
  \label{tbl:cut_running_times}
  \end{center}
\end{table}

\section{Conclusions \& Future Work}\label{sec:conclusions_future_work2}

We have presented an algorithm that allows a user to perform 
unconstrained consecutive tears on a rigged model in VR, while 
preserving its ability to be deformed as a soft-body. 
Since our method is geometry-based, it does not require 
significant GPU/CPU resources, it 
is amenable to work in real-time VR even for low-spec devices, making 
it ideal for mobile VR. We expect that 
it will eventually pave the way to alter the modern landscape of such VR interactions, where similar operations are mostly predefined. 
Also, most state-of-the-art methods including physically-correct methods (e.g. Finite Element Methods) cannot be used as they require significant computing resources and/or produce low fps results, unsuitable for mobile VR applications.
The proposed framework is already implemented in the MAGES SDK, 
running on Unity3D, publicly available for free.

In the future, we intend to further optimize our 
framework to work in a fraction of the current running times by  
utilizing  GPU compute and geometry shaders, taking advantage of 
the parallel pipeline they offer. So far, the 
performance overhead by our framework 
during a VR session, involving high-complexity models, 
is minimal and in most cases
negligible, due to the user's mental preparation time between actions. 
As tear operations are especially useful for VR medical training 
scenarios, we would like to explore our algorithm's 
adaptation to the collaborative needs of multi-user scenarios 
of such applications. 
Lastly, we would like to investigate the utilization of deep learning for the optimal 
identification of best suited clusterings, based on the model,
and the action(s) the user intents to perform. 

\section*{Acknowledgments}
The project was partially funded by the European Union’s 
Horizon 2020 research and innovation programme under grant agreements 
No 871793 (ACCORDION) and No 101016509 (CHARITY).

\bibliographystyle{ACM-Reference-Format}
\bibliography{references.bib}

\end{document}